%% file: PRD_O3_ASAF.tex
\documentclass[aps, prd, showpacs, preprintnumbers, floatfix,longbibliography,superscriptaddress,reprint]{revtex4-1}

\usepackage{graphicx,url,amssymb,amsmath,longtable,rotating,color,wasysym,subfigure,epsfig,epstopdf}

\usepackage[colorlinks=true, linkcolor=blue, citecolor=teal]{hyperref}
\usepackage[dvipsnames]{xcolor}

\usepackage{enumitem}
\usepackage{appendix}

\usepackage{graphicx}
\usepackage{dcolumn}
\usepackage{bm}
\usepackage{url}
\usepackage{array,multirow}
\usepackage{color} 
\usepackage{aas_macros} 
\usepackage{ifthen}
\usepackage{etoolbox}
\usepackage{physics}
\usepackage{siunitx} 
\usepackage{acronym,cleveref}
\usepackage{soul}
\newtoggle{fullauthorlist}
\toggletrue{fullauthorlist}

\newtoggle{endauthorlist}
\togglefalse{endauthorlist}
\usepackage{xr}

\acrodef{LIGO}{Laser Interferometer Gravitational-wave Observatory}
\acrodef{aLIGO}{Advanced LIGO}
\acrodef{GW}{gravitational waves}
\acrodef{CGB}{cosmological gravitational-wave background}
\acrodef{SNR}{signal-to-noise ratio}

\begin{document}

\title{All-sky, all-frequency directional search for persistent gravitational-waves from Advanced LIGO's and Advanced Virgo's first three observing runs}
\date{\today}

\iftoggle{endauthorlist}{
  \let\mymaketitle\maketitle
  \let\myauthor\author
  \let\myaffiliation\affiliation
  \author{The LIGO Scientific Collaboration}
  \author{The Virgo Collaboration}
  \author{The KAGRA Collaboration}
  \email{Full author list given at the end of the article.}
\noaffiliation
}{
  \iftoggle{fullauthorlist}{
  \input{LSC-Virgo-KAGRA-Authors-Feb-2021-prd}
  }{
    \author{The LIGO Scientific Collaboration}
    \affiliation{LSC}
    \author{The Virgo Collaboration}
    \affiliation{Virgo}
    \author{The KAGRA collaboration}
    \affiliation{KAGRA}
  }
}

\begin{abstract}

We present the first results from an all-sky all-frequency (ASAF) search for an anisotropic stochastic gravitational-wave background using the data from the first three observing runs of the Advanced LIGO and Advanced Virgo detectors. Upper limit maps on broadband anisotropies of a persistent stochastic background were published for all observing runs of the LIGO-Virgo detectors. However, a broadband analysis is likely to miss narrowband signals as the signal-to-noise ratio of a narrowband signal can be significantly reduced when combined with detector output from other frequencies. Data folding and the computationally efficient analysis pipeline, {\tt PyStoch}, enable us to perform the radiometer map-making at every frequency bin. We perform the search at 3072 {\tt{HEALPix}} equal area pixels uniformly tiling the sky and in every frequency bin of width $1/32$~Hz in the range $20-1726$~Hz, except for bins that are likely to contain instrumental artefacts and hence are notched. We do not find any statistically significant evidence for the existence of narrowband gravitational-wave signals in the analyzed frequency bins. Therefore, we place $95\%$ confidence upper limits on the gravitational-wave strain for each pixel-frequency pair, the limits are in the range $(0.030 - 9.6) \times10^{-24}$. In addition, we outline a method to identify candidate pixel-frequency pairs that could be followed up by a more sensitive (and potentially computationally expensive) search, e.g., a matched-filtering-based analysis, to look for fainter nearly monochromatic coherent signals. The ASAF analysis is inherently independent of models describing any spectral or spatial distribution of power. We demonstrate that the ASAF results can be appropriately combined over frequencies and sky directions to successfully recover the broadband directional and isotropic results. 
\end{abstract}

\maketitle

\section{Introduction}
A stochastic gravitational-wave background (SGWB)~\cite{AllenOttewill,Romano2017} can be created by the superposition of a large number of unresolved independent sources~\cite{2018PhRvL.120i1101A,2011PhRvD..84h4004R, 2011PhRvD..84l4037M, 2011ApJ73986Z, 2012PhRvD..85j4024W, 2013MNRAS.431..882Z,2016PhRvD..94j3011D, 2021PhRvD.103d3002P,hotspot,HUGHES201486,2018arXiv180710620C,PhysRevD.88.062005,PhysRevD.50.1157,PhysRevD.85.023534,CROWDER201366,Regimbau_review,Inayoshi,Watanabe,Kamionkowski,Kosowsky,Turner}. Improvements in detector sensitivity suggest that the network of ground-based gravitational-wave (GW) observatories may be able to observe such a background in the coming years. Astrophysical sources in the nearby universe can make the background anisotropic \cite{PhysRevD.98.063509,2014PhRvD..89h4076M,PhysRevLett.120.231101, 2018PhRvD..98f3501J,2012PhRvD..86j4007R, 2013PhRvD..87d2002W,2013PhRvD..87f3004L,PhysRevD.96.103019}. Directional searches were performed~\cite{O1paper,O2paper,Renzini_aniso,O1O2Folded,o3_direction} on data from the Advanced LIGO-Virgo detectors for two different source categories to probe these anisotropies.
One search focuses on a persistent SGWB from a collection of point-like and extended distributions of sources emitting GWs over a broad frequency range~\cite{radio_method,Mitra07,Eric09,pystochSpH,o3_direction}. The second search looks for narrowband signals from known locations of potentially detectable continuous wave sources~\cite{o3_direction}.

The past analyses, however, had limited prospects of detecting an unknown narrowband anisotropic stochastic background.
The broadband searches are not optimised to detect narrowband signals.
While the broadband radiometer search~\cite{Mitra07,radiometer} is capable of coherently adding signals from multiple narrowband sources, noise from thousands of other frequency bins that do not contain any signal can degrade the signal-to-noise ratio.
Moreover, it is straightforward to combine results from a narrowband analysis with appropriate frequency-dependent weight factors to derive the broadband results for a variety of spectral shapes including non-power law predictions in astrophysical ~\cite{Regimbau_review,Inayoshi,Marassi_CT} and cosmological~\cite{LVK_CS,Christensen_sgwb,Binetruy_PT_CS} scenarios.
Such efforts can be further extended to allow the spectral shape to vary across the sky, which is expected when multiple anisotropic backgrounds are simultaneously present in the sensitive frequency bands of the detectors.
Performing the directional search at all the narrowband frequency bins separately is thus well motivated. Recent developments on data folding~\cite{folding} and the introduction of a new analysis pipeline {\tt{PyStoch}}~\cite{pystoch}, that have made the radiometer analysis hundreds of times faster, opens up the possibility of performing an all-sky directed radiometer search for unknown persistent signals in narrow frequency bins~\cite{eric_coarse_grain,boris}. 

In addition, many unknown galactic and extra-galactic continuous GW sources, primarily from neutron stars~\cite{lorimer,cw-review,LIGOScientific:2019lyx,Zhang:2020rph,Espinoza2011ASO,PhysRevD.96.063004,PhysRevD.77.082001,PhysRevD.94.082004,LIGOScientific:2008hqb,PhysRevD.99.084048,PhysRevD.101.082004} and exotic scenarios like boson clouds around spinning black holes~\cite{Brito_Boson,leo_boson}, may be detectable with the Advanced LIGO-Virgo detectors.
Searching for such signals from neutron stars across the sky using known models is always computationally expensive~\cite{o3-cw-allsky,Ming:2021xtz}. Since these matched-filtering-based searches depend on the signal model via frequency evolution templates, the chances of detecting an unknown or poorly modeled family of sources (e.g., accreting and/or short-period binary systems) are limited and expanding the parameter space is computationally challenging~\cite{cw-review,PhysRevD.58.063001}. Also, the probability of detection in such matched-filtering-based searches is reduced substantially in the presence of glitches~\cite{cw-glitch}. The all-sky all-frequency (ASAF) analysis is robust with respect to modeling and able
to rapidly identify candidate frequency - sky location pairs,
that may warrant following up with more sensitive matched filtering-based searches.

The ASAF results could also be used to look for cross-correlations between SGWB and the electromagnetic sky~\cite{Suvodip,Yang_gwem,Adshead:2020bji,Malhotra:2020ket}, where each narrowband frequency bins may be analysed independently for setting stronger constraints on astrophysical and cosmological models.

In this paper we report the first ASAF upper limits on an unmodeled anisotropic SGWB using data from the first three observing runs (O1+O2+O3) of the Advanced LIGO and Advanced Virgo detectors. 

\section{Data}
To perform the ASAF search, we analyze the data from the first (O1) and second (O2) observing runs of the Advanced LIGO~\cite{2015CQGra..32g4001L} detectors located in Hanford (H) and Livingston (L), and from the third observing runs (O3) of both Advanced LIGO and Advanced Virgo~\cite{2015CQGra..32b4001A} (V). The recorded data have been processed and conditioned in the same way as was done in the latest directional search analysis by the LIGO-Virgo-KAGRA (LVK) collaboration ~\cite{o3_direction,cw_dq,gatingDocument,stochasticGatingDocument}. To account for non-Gaussian features in the data, we identify and remove segments containing detected GW signals from compact binary coalescences~\cite{gwtc-1,gwtc-2,o3_direction}, non-retracted events in the second half of O3~\cite{gracedb}, transient hardware injections (simulated signals injected by physically displacing the interferometer mirrors~\cite{CW_hardware_injection}) and segments associated with instrumental artifacts.
We then obtain the cross-spectral density (CSD) by combining the short-term Fourier transforms (SFTs) of $192$ second data segments from pairs of detectors~\cite{allen-romano}, and the corresponding variances, with a coarse-grained frequency resolution of $\Delta f = 1/32$ Hz~\cite{o1_directional}.

We apply the ASAF analysis in the frequency range 20-1726 Hz as in the past stochastic searches. In addition to the time-domain data quality cuts, we also identify contaminated frequency bins using coherence studies~\cite{o3_iso}. These frequency bins are mainly associated with known instrumental artifacts (calibration lines, power lines, and their harmonics). We also remove the continuous-wave hardware injections~\cite{CW_hardware_injection}  in the final analysis, though we use these for validating the analysis pipeline.
Since we produce results for each frequency bin separately, it is important to have a stringent check on the noise contamination for all the individual bins. In the broadband search~\cite{o3_direction}, while all attempts are made to discard frequency bins with noise contamination, to avoid too much loss of sensitivity, notching is relatively less stringent, as noise in a few frequency bins has an insignificant effect on the whole integrated frequency band.
We apply a more stringent threshold on the permissible level of non-stationarity in individual frequency bins compared to the broadband search~\cite{asaf_supplementary}.
This choice results in removing approximately 21\%, 34.8\%, and 28.2\% of the frequency bins from O3 data for HL, HV, and LV baselines, respectively, compared to 14.8\%, 25.2\%, 21.9\%, for the non-stringent notching used in the broadband search~\cite{o3_direction}.
After combining the baselines and the three observing runs, the number of completely notched frequencies reduces to 12.5\% of the total.

\section{ASAF Estimators}
In contrast to the previous stochastic directional analyses, which constrained either broadband integrated anisotropy of the sky or narrowband sources in specific directions, the ASAF analysis attempts to probe anisotropy of the whole sky in each frequency bin.
The anisotropies can be characterized in terms of a dimensionless GW energy density parameter $\Omega_{\rm GW}(f,\hat{\mathbf{n}})$ with units sr$^{-1}$, defined as
~\cite{allen_1997,allen-romano,o3_direction},
\begin{equation}\label{eq:Omega_GW}
    \Omega_{\rm GW}(f,\hat{\mathbf{n}})\ \equiv \ \frac{f}{\rho_{\rm c}}\frac{\dd \rho_{\rm GW}}{\dd f} \ =  \ \frac{2\pi^2}{3H^2_0}f^3 \mathcal{P}(f,\hat{\mathbf{n}})\ ,
\end{equation}
where $\rho_{\rm GW}$ is the energy density of incoming GWs from the direction $\hat{\mathbf{n}}$ observed in the frequency range from $f$ to $f+df$, $\rho_{\rm c} = 3 c^2 H_0^2/8 \pi G$ is the critical energy density needed to close the universe, $c$ denotes the speed of light, $G$ is the  gravitational constant, and $H_0$ is the Hubble constant. We are interested in estimating the sky-maps $\mathcal{P}(f,\hat{\mathbf{n}})$ at every frequency, which can be decomposed as, 
\begin{equation}
    \mathcal{P}(f,\hat{\mathbf{n}}) \ = \ \sum_p \ \mathcal{P}_p(f)\, e_p(\hat{\mathbf{n}})\ ,
\end{equation}
in terms of basis functions $e_p(\hat{\mathbf{n}})$. The choice of the basis usually depends on the target source. For extended sources, the spherical harmonic basis $Y_{lm}(\hat{\mathbf{n}})$ is a common choice, while for localised point-like sources the pixel basis, $e_p(\hat{\mathbf{n}}) = \delta^2(\hat{\mathbf{n}}-\hat{\mathbf{n}}_p)$, where $\hat{\mathbf{n}}_p$ is the direction of pixel number $p$, is an appropriate choice. Here we perform the ASAF analysis in pixel basis only.

To measure the anisotropy $\mathcal{P}(f,\hat{\mathbf{n}})$, the radiometer search uses the maximum likelihood (ML) estimator~\cite{sph_methods,Mitra07} 
as the statistic,
\begin{equation}
    \bm{\hat{\mathcal{P}}} (f)  \ = \ {\mathbf{\Gamma} (f)}^{-1}\,\mathbf{X} (f)\, ,
    \label{eq:estimator}
\end{equation}
where $\mathbf{X}$ is the \textit{dirty map}~\cite{Mitra07} and $\mathbf{\Gamma}$ is the Fisher information matrix~\cite{sph_methods} in the weak signal limit, where in the chosen basis, the signal is much smaller than the standard deviation of the noise in each data segment. The dirty map $\mathbf{X}$ represents the SGWB sky seen through the response matrices of a baseline $\mathcal{I}$ formed by the detectors $\mathcal{I}_1$ and $\mathcal{I}_2$, defined as
\begin{equation}
X_p(f) \ = \ \tau \Delta f \,  \mathfrak{Re} \sum_{\mathcal{I},t}  \frac{ \gamma^{\mathcal{I}*}_{ft,p}\,C^{\mathcal{I}} (t;f)} {P_{\mathcal{I}_1}(t;f)\, P_{\mathcal{I}_2}(t;f)}  \,, 
\label{eq:Dirty_map}
\end{equation}
where 
$\tau$ is the length (duration) of each time segment,
$P_{\mathcal{I}_{1,2}}(t;f)$ denotes the one-sided noise power spectral density (PSD) and $C^{\mathcal{I}} (t;f) := (2/\tau) \, \widetilde{s}^*_{\mathcal{I}_1}(t;f) \, \widetilde{s}_{\mathcal{I}_2} (t;f)$ is the CSD, $\widetilde{s}_{\mathcal{I}_{1,2}}(t;f)$ are the SFTs of data from the detectors at a time-segment marked by $t$. 
In practice, to prevent spectral leakage without loss of data, overlapping Hanning windows are applied to the time series data in each segment, introducing additional normalisation factors, which have been accounted for in the analysis~\cite{albert_joe,RomanoCornish,radio_method,radiometer,o1_directional}. Also, the coarse-grained frequency bin size $\Delta f$ is greater than $1/\tau$, i.e., $\tau \Delta f$ is not unity~\cite{Andrew_Joe,o3_iso}.
The overlap reduction function (ORF), $\gamma^{\mathcal{I}}_{ft,p}$, is defined as
\begin{equation}
\gamma^{\mathcal{I}}_{ft,p}:\equiv \,\sum_{A}  F^{A}_{\mathcal{I}_1}(\mathbf{\hat {n}}_p,t) \,
F^{A}_{\mathcal{I}_2}(\mathbf{\hat{n}}_p,t) \, e^{2\pi i f\, \mathbf{\hat{n}}_p\cdot {\mathbf{\Delta x}_{\mathcal{I}} (t)}/c},
\label{eq:orf}
\end{equation}
where ${\mathbf{\Delta x}_{\mathcal{I}} (t)}$ is the separation vector between the detectors. In the above equation $A=+,\times$ denotes the polarization (note that this analysis assumes statistically equivalent $+$ and $\times$ polarizations). The ORF is necessary to optimally ``point'' the radiometer~\cite{Mitra07} to a direction $\mathbf{\hat{n}}_p$ corresponding to the pixel index $p$ by cross-correlating data streams from pairs of detectors with time varying phase delay, along with the sky modulation induced by the antenna pattern functions of the detector ($F^A_{\mathcal{I}_{1,2}}$)~\cite{allen-romano,Romano2017}. The uncertainty in the dirty map measurement is encoded in the Fisher information matrix defined as
\begin{equation}
\Gamma_{pp'} (f) \ \equiv \ \frac{\tau\Delta f}{2} \, \mathfrak{Re} \sum_{\mathcal{I},t} \frac{\gamma^{\mathcal{I}*}_{ft,p} \, \gamma^{\mathcal{I}}_{ft,p'}}{P_{\mathcal{I}_1}(t;f) \, P_{\mathcal{I}_2}(t;f)} \, .
\label{eq:Fisher}
\end{equation}
Since the estimators are obtained by summing over a large number of $\tau = 192\,\mathrm{second}$ time-segments, the Central Limit Theorem implies that the noise distribution
is approximately Gaussian as long as the total observation is longer than a few hours.

\begin{figure*}
    \centering
    \includegraphics[height = 0.25\textwidth]{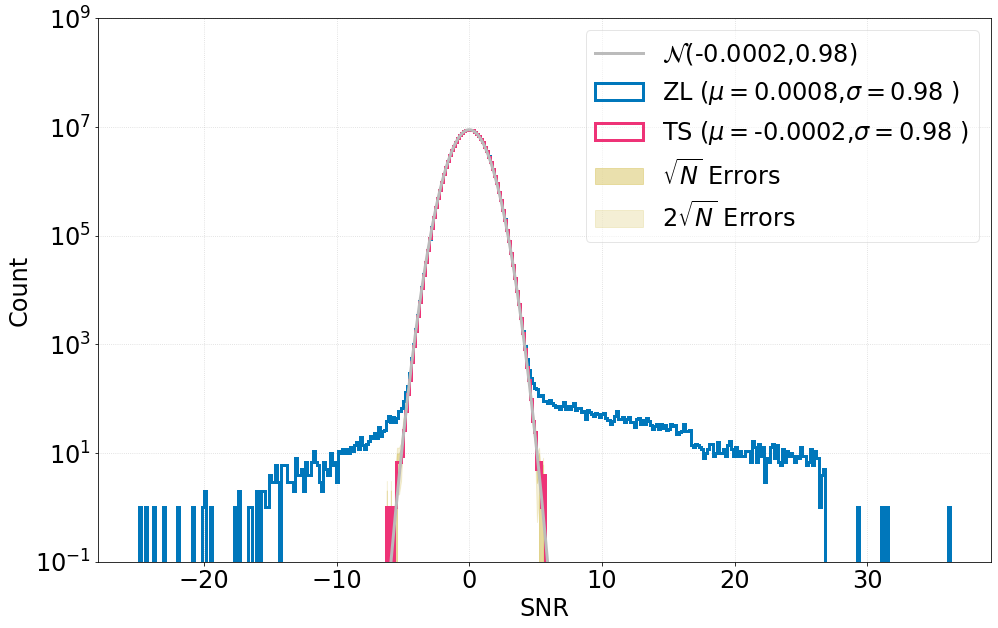}
    \includegraphics[height = 0.25\textwidth]{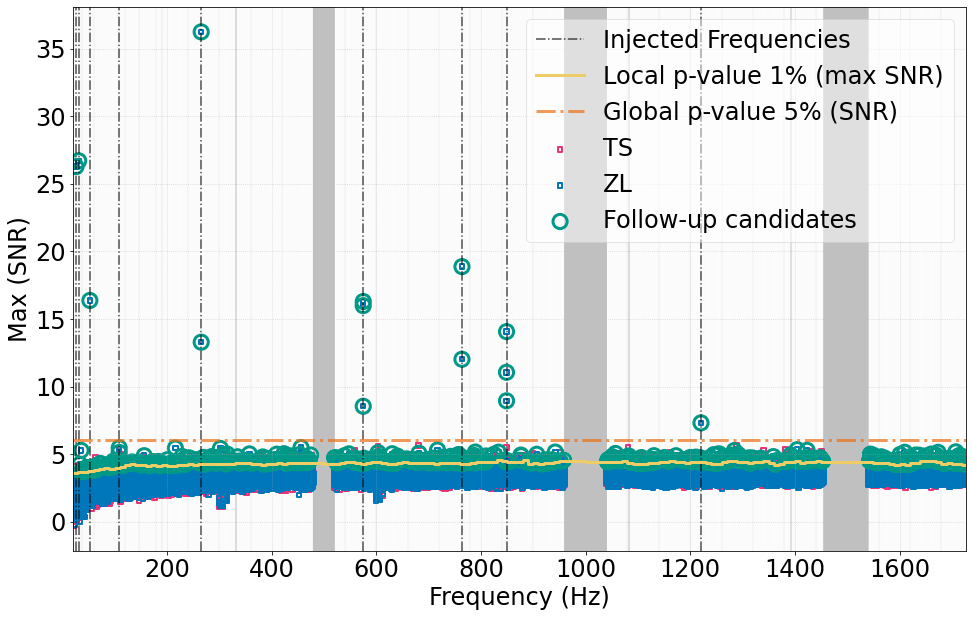}\\
    \includegraphics[height = 0.25\textwidth]{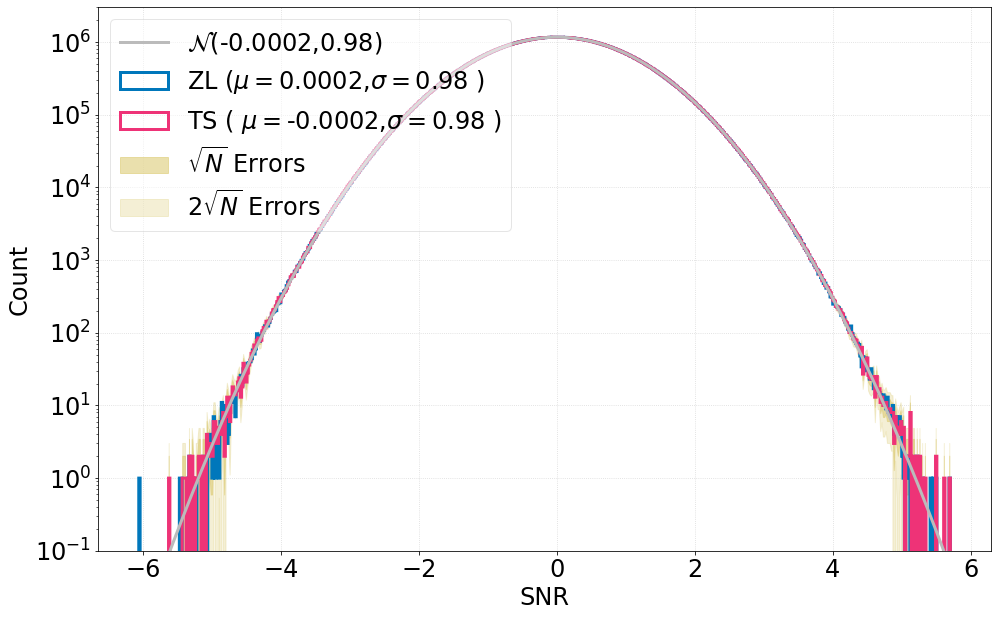}
    \includegraphics[height = 0.25\textwidth]{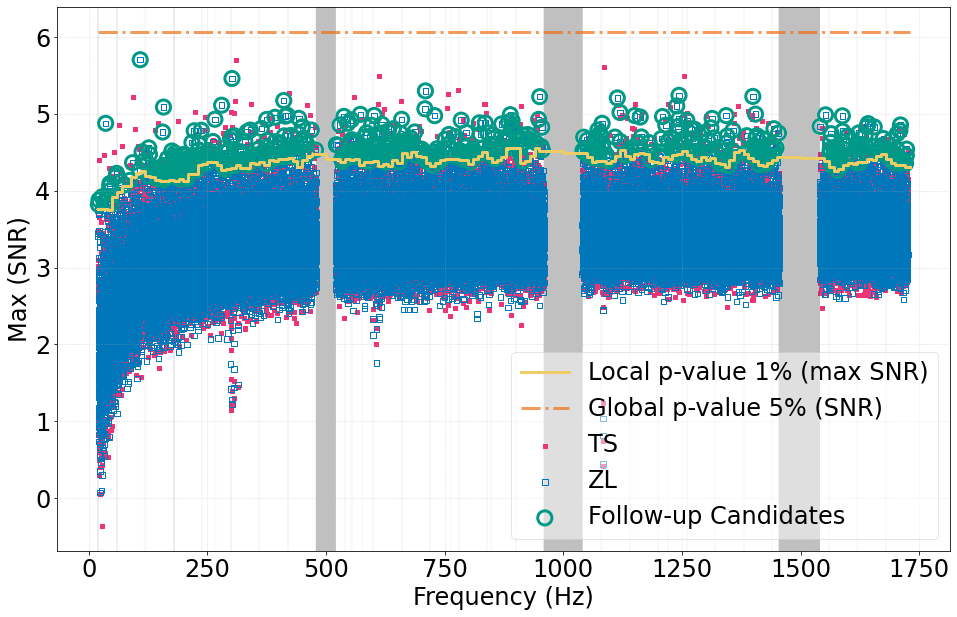}
    \caption{Results from the O2 hardware injection study (top panel) and O1+O2+O3 data sets (bottom panel) are shown.  Left hand side plots  in the top and bottom panels show the histogram of SNR from both the (unphysical) time-shifted  (TS)  analysis (red) and zero-lag  (ZL) analysis (blue). Time-shifted and zero-lag data sets are both consistent with Gaussian noise (grey histogram) within the $2\sigma$ error bars. The outliers cause the extended tails in blue histogram representing the distribution of foreground SNRs. These extended tails in the top left plot depicts the recovery of the hardware injections performed on the O2 data. The point with SNR $\sim -6$ in bottom left plot is not astrophysically motivated and most likely caused by low number statistics (the two-tailed $p$-value is more than 5\%). The right hand side plots in the top and bottom panels show the distribution of maximum SNRs from both the time-shifted and zero-lag analyses. In these plots, the grey solid vertical lines represent the frequencies notched in the data, the yellow curve shows the 99th percentile of maximum SNR for every 10 Hz frequency bin in the time-shifted analysis, smoothed over $3$ neighboring 10 Hz bins, and the orange dot-dashed line delineates the trials-factor-corrected, one-sided global $p$-value of $5\%$. The points above the yellow curve, the identified candidates for follow-up studies, are marked with teal circles. For the O2 hardware injection study, one can see that (top right) all the injections are recovered which appear as multiple teal circles very close to the black vertical lines indicating the frequency bin where the injections were made. On the other hand, for O1+O2+O3 data set, we do not find any outliers significantly above the noise background (bottom right).  In this case, the teal circles represent $515$ candidates which may be followed up by a more sensitive matched-filtering-based analysis.}
    \label{fig:inj}
\end{figure*}

Once $X_p(f)$ and $\Gamma_{pp'} (f)$ are calculated using Eqs.~\ref{eq:Dirty_map} and \ref{eq:Fisher} for all the baselines and observing runs, we combine them to obtain the multi-baseline (HLV here onward) dirty map and Fisher information matrix for all the observing runs. From the combined Fisher matrix and dirty map, we can construct the estimator in Eq.~\ref{eq:estimator}. The ML estimator $\bm{\hat{\mathcal{P}}}$ involves inversion of the Fisher matrix $\mathbf{\Gamma}$ which has singular values associated with certain observed modes on the sky where the baselines are insensitive. For point sources, given the current sensitivity of detectors and the pixel resolution used here, correcting for the pixel-to-pixel correlation hardly makes any difference to the analysis~\cite{O1O2Folded}. Therefore, to obtain the estimator of narrowband anisotropy and the corresponding signal-to-noise ratio (SNR), instead of inverting the Fisher matrix we divide the dirty map in a given pixel by the corresponding diagonal element of the Fisher matrix. Then the estimator of narrowband anisotropy and its uncertainty are given by,
\begin{eqnarray}
    \hat{\mathcal{P}}(f,\hat{\mathbf{n}})&=&[{\Gamma_{\hat{\mathbf{n}},\hat{\mathbf{n}}}(f)}]^{-1} \, X_{\hat{\mathbf{n}}}\,,\\
    \sigma_{\hat{\mathbf{n}}}(f)&=&[{\Gamma_{\hat{\mathbf{n}},\hat{\mathbf{n}}}(f)}]^{-1/2}\,.
\end{eqnarray}
Now one can write the observed (SNR) as
\begin{equation}
    \rho(f,\hat{\mathbf{n}})=\hat{\mathcal{P}}(f,\hat{\mathbf{n}})/\sigma_{\hat{\mathbf{n}}}(f) \,.
\end{equation}
In the absence of any signal, $\rho(f,\hat{\mathbf{n}})$ follows a Gaussian distribution (Fig.~\ref{fig:inj}). This formalism is used to perform the ASAF analysis.

\section{Data folding and {\tt{PyStoch}}}
Folding~\cite{folding} makes use of the temporal symmetry in the detector scan pattern to compress the data into a single sidereal day. This reduces the computational cost of the search by a factor equal to the total number of days in the observation run.

{\tt{PyStoch}}~\cite{pystoch} is a fully python-based standalone pipeline for SGWB map-making that takes full advantage of the compressed folded data and symmetries in the detector set-up, further improving the efficiency of the search with respect to the previous analysis pipeline.
Both folding and {\tt{PyStoch}} were recently adapted for the broadband directional search in LIGO-Virgo data~\cite{o3_direction}.

Since folding ensures that
the data size is fixed (one sidereal day long)
and can be loaded entirely in the memory of most computers, {\tt{PyStoch}} is able to cast the
segment-wise radiometer analysis
to a matrix multiplication problem incorporating all the time segments together. {\em This allows interchangeable ordering of operations over time and frequency, which is essential for the ASAF analysis}, and efficient parallel processing of data. Here we use {\tt {HEALPix}}~\cite{HEALPix} resolution of $N_{\mathrm{side}} = 16$, which corresponds to the number of pixels $N_{\mathrm{pix}} = 12 N_{\mathrm{side}}^2 = 3072 $. In principle, one could use a lower (higher) pixel resolution at lower (higher) frequencies. Since the chosen resolution is adequate for the most sensitive frequency band of the baselines, we refrain from introducing further complexities or using higher resolution in the present analysis.
With these search parameters, it took less than an hour per baseline for each data set to run on a personal computer~\cite{laptop}.

\section{Detection statistics and Outliers}
In order to search for significant outliers, which could indicate the presence of a signal, we first need to determine the noise background. Since it is impossible to directly measure the detector noise in the absence of persistent GW signals, the background is estimated by introducing a constant unphysical \textit {time-shift}~\cite{o3_direction,o3_iso} of $\sim 1$~second between the data streams from a pair of detectors which is much greater than the light travel time between the interferometers. If a signal is not correlated on longer time scales, which is the case for black hole mergers or the stochastic background created by such ``popcorn'' type events~\cite{Regimbau_2011}, these time-shifts are expected to remove the signal correlation between detectors and the resultant distribution of the detection statistic will then represent the noise background~\cite{S5_Dir}. However, these constant time-shifts do not completely cancel out the types of sources that have longer coherence. For example, if we consider an isolated neutron star emitting GWs at a certain frequency, the constant time-shift analysis will not remove the presence of such a signal; the signal might appear to originate from a different direction or with an unphysical negative SNR.
Since the ASAF results could be useful for identifying potential locations and frequencies of previously unknown neutron stars, we apply an alternative method to alleviate this problem.

We estimate the noise background by adding a \textit{random} time-shift to each contiguous slice of data such that the net shift (1-2~second) is always much greater than the physically permissible time delay. In order to validate this prescription, we use continuous hardware injections of isolated pulsars with varying signal strength performed in Advanced LIGO's second (O2) observation run (the hardware injections for O3 were too weak~\cite{o3-cw-allsky} for the ASAF analysis to recover with enough SNR). To generate the background we first apply the random time-shift for the Hanford-Livingston pair and run the entire analysis. These SNRs are plotted in the histogram shown in Fig.~\ref{fig:inj} along with a Gaussian fit. 
It is clear that the time-shifted method is successful in removing the effect of the injected signals (the elevated tails of the distribution), leading to a Gaussian noise background. 
Had the ``foreground'' distribution---the distribution of SNR without any unphysical time-shifts (zero-lag)---been inside the $2\sigma$ error bars around the Gaussian, we would rule out the presence of any outliers with $95\%$ confidence. 
The presence of some large negative SNR values for zero-lag in Fig.~\ref{fig:inj} is due to the mismatch between the circular polarization model implicitly assumed in Eq.~\ref{eq:orf} for a continuous-wave source
and the elliptic polarizations simulated by the hardware injections, some of which are nearly linearly polarized.
We searched for 9 injections that were in the analysed frequency range and with 
source-frame frequency variation
less than $1/32$~Hz~\cite{o2_HWI}. Among these, 8 were detected as ``outliers'' and one as a follow-up candidate (the ninth injection is slightly below the detection threshold, but above the threshold for identifying as a follow-up candidate). The maximum dirty map SNRs are plotted in Fig.~\ref{fig:inj}.
In some cases, an injection is recovered also in 
the previous or the next frequency bin
due to spectral leakage~\cite{RomanoCornish},
which appears as multiple circles very close to one vertical line. The recovered locations of the sources match the locations of injections within the diffraction-limited resolution.
Note that, as anticipated, the broadband search did not detect any outliers~\cite{asaf_supplementary}.

In order to identify potential candidates for follow-up, we consider the distribution of maximum pixel SNR, $\rho_{{\rm max}}(f)\equiv{\rm max}_p\,[\rho(f,\hat{\mathbf{n}}_p)]$. The maximum SNR are shown as a scatter plot in the right panel of Fig.~\ref{fig:inj}. We divide the whole frequency range into 10 Hz bins, over which the sensitivity of the radiometer does not vary significantly. We make a histogram of $\rho_{\rm{max}}(f)$ for all the 1/32 Hz frequency bins in each 10 Hz bin, separately for zero-lag and time-shifted analysis. For each 10 Hz bin, we find the $\rho_{\rm{max}}(f)$ that corresponds to 99th percentile ($1\%$ false alarm rate) of the maximum SNR distribution for time-shifted data and average it over $3$ neighboring 10 Hz bins (yellow line in the right panel of Fig.~\ref{fig:inj}). Any $\rho_{\rm{max}}(f)$ that is above this threshold in the zero-lag analysis is marked as a candidate for follow-up studies.
All hardware injections lie above this threshold curve and qualify for detailed follow-up studies. The ``slope'' of the maximum SNR distribution in the scatter plot changes because the diffraction-limited resolution is a function of frequency~\cite{boris,Mitra07}. Therefore, at lower frequencies, 
the number of independent sky patches is smaller and,
hence, the pixels are more correlated. The correlation reduces at higher frequencies.

Reassured by the above hardware injection study, we apply the same procedure to obtain the significance of the results from the ASAF search. The distribution of SNR at each frequency and pixel on the sky are shown in the histogram in the bottom left panel of Fig.~\ref{fig:inj} along with the SNR distributions from the time-shifted run. It is evident from the figure that the distribution of SNRs follow the noise background and in turn is consistent with Gaussian noise within $2\sigma$ error bars. Since the number of frequency-pixel pairs is $\sim 10^8$, there is greater than a $5\%$ probability of at least one high SNR ($\sim 6$) observation arising purely from noise, as seen in the tails of the distributions.
{\em The ASAF analysis thus rules out the existence of any significant outliers in the O1+O2+O3 data}.

We nevertheless apply the same procedure described for hardware injections to identify potential candidates that can be followed up by a more sensitive search. We again consider the distribution of maximum SNRs in each 10 Hz frequency bin. These maxima are plotted in the bottom right panel of Fig.~\ref{fig:inj}. Here, the noise background obtained from the time-shifted method is shown in red in the scatter plot whereas the results from zero-lag data are represented in blue. The yellow line delimits the SNRs above the 99th percentile of the background obtained from the unphysical time-shifted data.
The maximum SNRs in the zero-lag analysis that pass this threshold can be classified as candidates for follow-up using template-based searches, even though they are not significant enough to be called outliers by the ASAF analysis---SNR is below the threshold of $6.05$ that corresponds to the trials-factor-corrected, one-sided global $p$-value of $5\%$ shown by the orange dot-dashed line.

\begin{figure*}
\centering
\includegraphics[width = 0.9\textwidth]{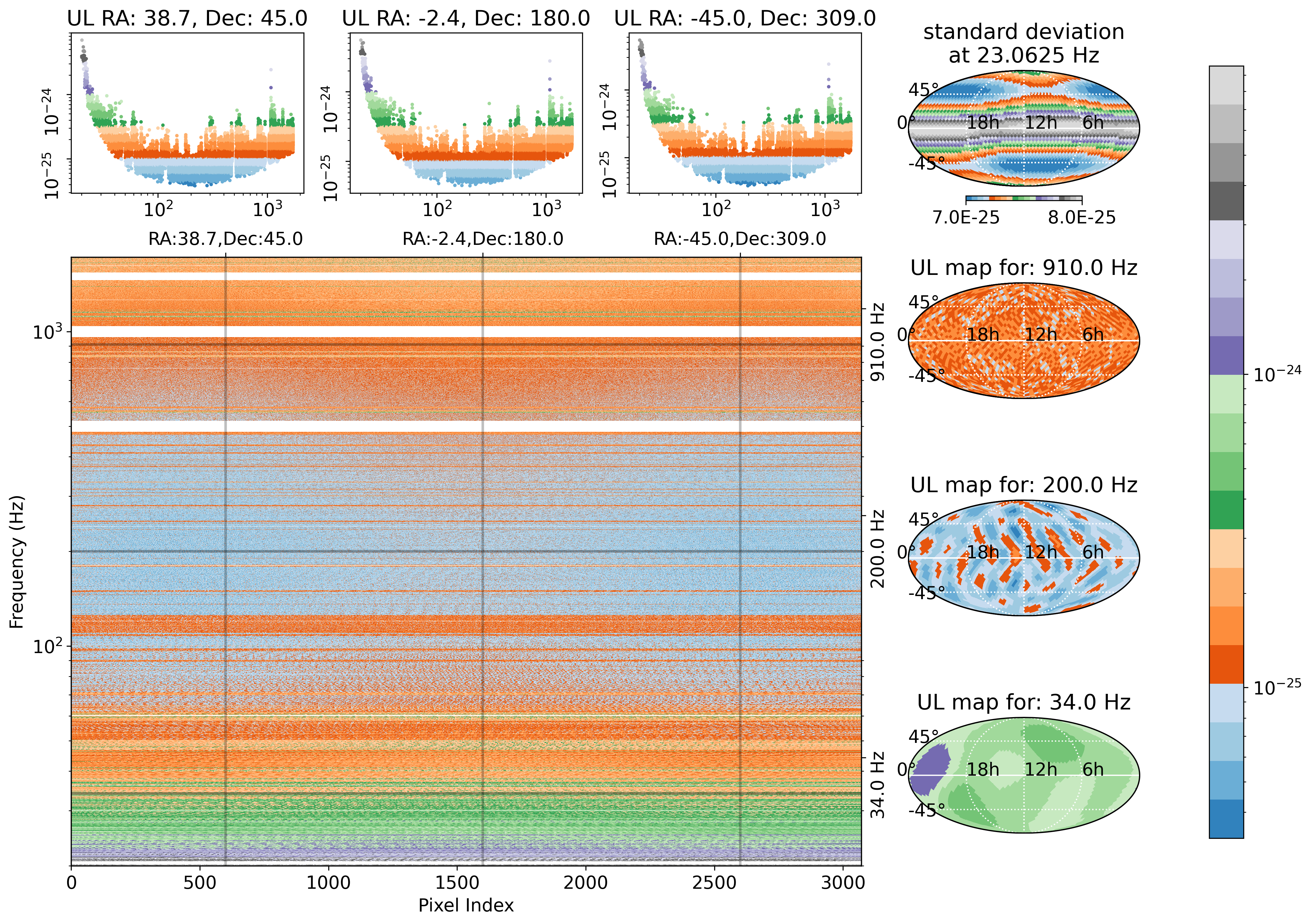}
\caption{The $95\%$ confidence Bayesian upper limit on the strain amplitude $h$ for all-sky directions and all-frequency bins. These upper limits are set using the ASAF search performed on the O1+O2+O3 data set. The color bar here denotes the range of upper limit variations. A vertical cross-section in this diagram shows the frequency dependent upper limit in a particular direction, 3 such plots are shown in the subplots on top. Whereas horizontal cross-sections form a map of upper limits in a particular frequency, 3 such maps are shown on the right. An example of the standard deviation in these maps is also shown in the top right corner.}
\label{fig:ASAF_Strain_UL}
\end{figure*}

\section{Upper limits}
In the presence of a detectable signal of strength $\mathcal{P}(f,\hat{\mathbf{n}})$, the ASAF point estimate, $\hat{\mathcal{P}}(f,\hat{\mathbf{n}})$, by construction would be distributed with a mean $\mathcal{P}(f,\hat{\mathbf{n}})$ and standard deviation $\sigma_{\hat{\mathbf{n}}}(f)$. Since the distribution is found to be consistent with the noise background obtained by unphysical time-shift analysis,
ruling out the detection of any significant signal,
we set upper limits on the strength of astrophysical signals. Here we take advantage of the fact that this noise background is Gaussian.
Since ASAF results are most relevant in the context of nearly monochromatic continuous wave signals, we set Bayesian upper limits on an equivalent strain amplitude of a circularly polarised signal, $h(f,\hat{\mathbf{n}})=\sqrt{{\mathcal{P}}(f,\hat{\mathbf{n}})\,df}$, without correcting for Doppler modulation caused by the earth's motion.
We assume a Gaussian likelihood for $\hat{\mathcal{P}}(f,\hat{\mathbf{n}})$ which is then marginalized over calibration uncertainty~\cite{Whelan_2014} with a uniform prior on $h(f,\hat{\mathbf{n}})$ in the wide range $0-10 \, \sigma$ of the point estimate. The $95\%$ confidence Bayesian upper limits placed on the HLV data set are shown in Fig.~\ref{fig:ASAF_Strain_UL}. The matrix plot shown in Fig.~\ref{fig:ASAF_Strain_UL} is a qualitative representation of the upper limit, where the upper limit sky maps are plotted as a function of {\tt HEALPix} pixel index on the horizontal axis and frequency on the vertical axis. The color bar represents the respective upper limit range. The horizontal ``bands'' and gaps in the matrix plot correspond to the notched frequencies in one or more detector pairs. These upper limits are in the range $(0.030 - 9.6) \times10^{-24}$. Note that, when interpreting upper limits on particular frequency-pixel pairs for potential point sources, such as neutron stars, allowance should be made for the angular distance between the source and the center of the pixel containing it; at the highest frequency searched (1726 Hz) the SNR for a point source at the edge of a pixel (at the chosen resolution) is $\sim 20$\% less than for the same source at the center of the pixel for most parts of the sky.
The inclusion of different baselines and observing run data reduces the fraction of completely notched frequencies and also improves the upper limits (more details are provided in~\cite{asaf_supplementary}).
In order to compare the relative sensitivities of different baselines and observing runs, one could define an effective strain sensitivity averaged over all pixels and frequencies (including the notched ones) as $[\langle \sigma^{-2}_{\hat{\mathbf{n}}}(f)\rangle/df^2 ]^{-1/4}$, which turns out to be $10^{-25}$ times 1.5, 0.89, 2.5, 2.2, 0.88 and 0.86 respectively, for O1+O2, HL, LV, HV \& HLV for O3 and HLV for O1+O2+O3.

We also include the plots of the upper limit sky maps for three example frequencies 34.0 Hz, 200.0 Hz and, 910.0 Hz in Fig.~\ref{fig:ASAF_Strain_UL}. The sky map on the top right corner shows how the uncertainty in the ASAF estimator is nearly axially symmetric and varies in latitude. This pattern is also reflected in the upper limit matrix plot.

\section{Derivation of broadband radiometer (BBR) estimates from ASAF search}
\begin{figure}
\centering
\includegraphics[width = \columnwidth]{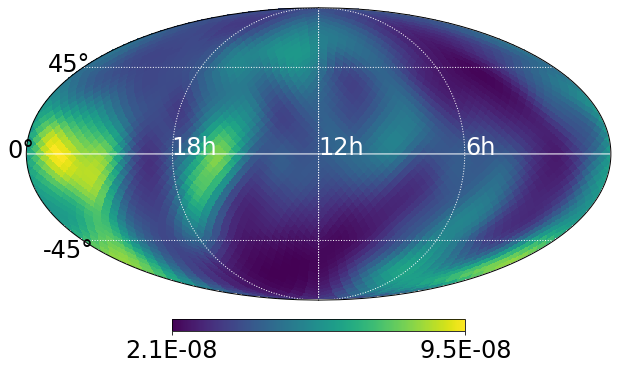}
\caption{The broadband $95\%$ confidence GW flux upper limit sky map for $\alpha = 0$ spectral shape. The sky map is represented as a color bar plot on a Mollweide projection of the sky in ecliptic coordinates.
}
\label{fig:bbr_spectra}
\end{figure}

One can integrate the ASAF results over frequency to obtain the broadband radiometer point estimate, variance and upper limits. We invoke the standard assumption used in the BBR search~\cite{o3_direction}, that the anisotropic PSD can be decomposed as a product of frequency and direction dependent terms,
\begin{equation}
    \mathcal{P}(f,\hat{\mathbf{n}}) =\mathcal{P}(\hat{\mathbf{n}})\, H(f) \,,
\end{equation}
where $H(f)$ is assumed to be a power law model proportional to $(f/f_{\rm{ref}})^{\alpha}$ where $\alpha$ is the spectral index and $f_{\rm{ref}}$ is a reference frequency set to $25~\rm{Hz}$. The BBR estimator and its standard deviation~\cite{S5dir,O1dir,O2dir} can then be written in terms of ASAF estimators as
\begin{eqnarray}\label{eq:BBR_estimators}
  \hat{\mathcal{P}}(\hat{\mathbf{n}})  =\frac{\sum_f \hat{\mathcal{P}}(f,\hat{\mathbf{n}})\, \sigma^{-2}_{\hat{\mathbf{n}}}(f)\,H(f)}{\sum_f\sigma^{-2}_{\hat{\mathbf{n}}}(f)\,H^2(f)}\,,\nonumber\\
  \sigma_{\hat{\mathbf{n}}}=\bigg[\sum_f\sigma^{-2}_{\hat{\mathbf{n}}}(f)\,H^2(f)\bigg]^{-1/2}\,.
\end{eqnarray}
Now one can calculate the GW flux in each direction $\hat{\mathbf{n}}$, 
\begin{equation}
\mathcal{F} (\mathbf{\hat{\mathbf{n}}}) \ = \ \frac{c^3 \pi}{4G}\,f_{\rm{ref}}^2\,\mathcal{P}(\mathbf{\hat{\mathbf{n}}})\, .
\label{eq:Flux}
\end{equation}
This quantity has the units of erg cm$^{-2} $s$^{-1}$ Hz$^{-1}$ sr$^{-1}$ measured with respect to the reference frequency. Using the methods outlined in~\cite{o3_direction}, we place upper limits on the amount of GW flux in each pixel. The upper limit range obtained from this analysis is reported in Table~\ref{tab:BBR_ISO_Results}. We have also shown in Fig.~\ref{fig:bbr_spectra} an example upper limit sky map derived from the ASAF search, performed on the O3 HL folded data for a specific spectral shape ($\alpha = 0$). These limits are consistent with the results we directly obtained from the broadband analysis of O3 data~\cite{o3_direction}. 

\section{Derivation of isotropic estimates from ASAF search}
From the ASAF search, it is also possible to derive the isotropic search results. One can integrate over all the sky directions to obtain the isotropic estimator~\cite{Ballmer_2006,S4_dir}. 
\begin{eqnarray}\label{eq:iso_estimator}
\hat{\mathcal{P}}_{\mathrm{\mathrm{iso}}}(f)\,\sigma^{-2}_{\mathrm{iso}}(f) = \frac{5}{4\,\pi} \int \, d\hat{\mathbf{n}}\, \hat{\mathcal{P}}(f,\hat{\mathbf{n}}) \,\sigma^{-2}_{\hat{\mathbf{n}}}(f)\,,\nonumber\\
    \sigma^{-2}_{\mathrm{iso}}(f) = \bigg(\frac{5}{4\,\pi}\bigg)^{2} \int \, d\hat{\mathbf{n}}\, \int \, d\hat{\mathbf{n}}'\,\Gamma_{\hat{\mathbf{n}},\hat{\mathbf{n}}'} (f)\,.
\end{eqnarray}
To compute the standard deviation of the isotropic estimator $\sigma_{\mathrm{iso}}(f)$, one requires the full narrowband pixel-to-pixel covariance matrix $\Gamma_{\hat{\mathbf{n}},\hat{\mathbf{n}}'}(f)$ which became computationally realistic with {\tt PyStoch} and folded data.
Conventionally isotropic searches report the SGWB in terms of $\Omega_{\mathrm{GW,iso}}(f)$. Eq.~\ref{eq:Omega_GW} can be used for the corresponding conversion of units. Since the {\tt HEALPix} grid is a discrete grid, we perform discrete integration with $d\hat{\mathbf{n}}=4\pi/N_{\mathrm{pix}}$.

\begin{figure}
\centering
\includegraphics[width = \columnwidth]{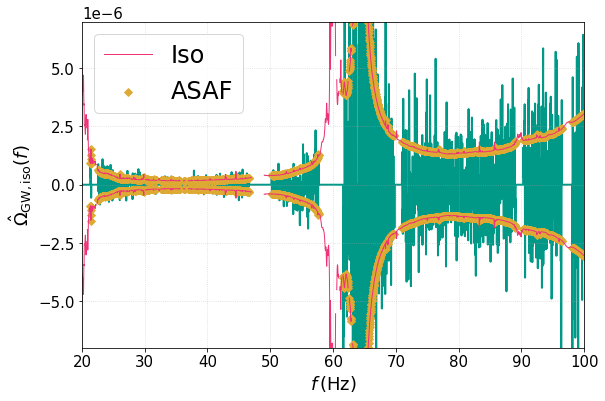}
\caption{The isotropic cross-correlation spectra derived from the ASAF analysis using the O3 HL baseline data. The red line depicts the estimated value of standard deviation with the isotropic ORF, while the yellow points show the uncertainty in the cross-power estimator obtained from the ASAF search. 
The green vertical line, which fluctuates around zero mean represents the point estimates from the ASAF search. The uncertainty calculated from the ASAF maps is not available at certain frequencies due to the more stringent notching.}
\label{fig:iso_spectra}
\end{figure}

The broadband isotropic estimators can then be derived using Eq.~\ref{eq:BBR_estimators} by replacing ASAF estimators with isotropic-all-frequency estimators given in Eq.~\ref{eq:iso_estimator}. The cross-correlation spectra from the O3 HL data set are shown (up to 100~Hz) in Fig.~\ref{fig:iso_spectra} and summarized in Table~\ref{tab:BBR_ISO_Results}.
The narrowband uncertainty obtained from the ASAF estimators and the one obtained using the analytically derived isotropic ORF are consistent up to few hundred Hz (which contributes more than 99\% sensitivity for the HL baseline~\cite{o3_iso}). 
On the other hand, the point estimate and standard deviation of the isotropic broadband estimate shown in Table~\ref{tab:BBR_ISO_Results} differ from our recently published results~\cite{o3_direction,o3_iso} due to different time and frequency domain data quality cuts and a different pixel resolution used in the ASAF analysis.

\begin{table}[b]
    \centering
    \begin{tabular}{c|c|c|c|c}
        \hline\hline
        \multicolumn{5}{c}{Broadband Results : HL baseline } \\
        \hline
       \hline
       \multicolumn{1}{c|}{$\alpha$}& \multicolumn{2}{c|}{Upper limit on $\mathcal{F} (\mathbf{\hat{\mathbf{n}}})$ ($\times10^{-8}$)} & \multicolumn{2}{c}{$\hat{\Omega}_\mathrm{GW,iso} (\times 10^{-9}$)}\\
        \hline
        \multicolumn{1}{c|}{}& ASAF & BBR~\cite{o3_direction} & ASAF & ISO~\cite{o3_iso}\\
         \hline
          0 & 2.1 - 9.6 & 1.8-9.2 & 2.5 $\pm$ 8.6  & -2.1 $\pm$ 8.2  \\
         \hline
         2/3 & 1.1 - 5.6 & 0.92-4.9 &  0.61 $\pm$ 6.5  & -3.4 $\pm$ 6.1\\
         \hline
          3 & 0.016 - 0.13 & 0.014-0.12  & -0.49 $\pm$ 0.99 & -1.3 $\pm$ 0.9\\
         \hline\hline
    \end{tabular}
    \caption{Broadband directional radiometer upper limit and the isotropic search results derived from the ASAF search (with the stricter notching) using the O3 HL data set. Results from the previous LVK O3 analyses are also added in the table for a direct comparison. The three spectral indices used in the search are denoted by $\alpha$.}
    \label{tab:BBR_ISO_Results}
\end{table}

\section{Conclusions}
We present the first all-sky all-frequency radiometer search results for data from ground-based laser interferometric detectors. No GW signal is detected by our analysis in the first three observing runs from the Advanced LIGO and Advanced Virgo detectors. We set 95\% confidence upper limits on the gravitational-wave strain at every frequency bin and sky location searched.

Note that, while the matched-filtering-based analyses can search for neutron stars at narrower frequency bins and are more sensitive~\cite{o3-cw-allsky} when such template-based searches are computationally feasible, ASAF analysis can rapidly search for such monochromatic signals with very little computation power and set upper limits at all frequency and sky-locations for the resolutions used here. The candidate frequency-pixel pairs identified by ASAF can be followed up by matched-filtering-based searches.
We employed a heuristic prescription for identifying these follow-up candidates. We recognize that alternative approaches may be explored in this regard. Future analyses will also examine the potential gains from refining frequency and pixel resolution. Since the total number of frequency-pixel pairs is very large ($\sim 10^8$), the prescription must be robust enough not to miss feasibly detectable signals, yet should limit follow-up candidates to a computationally viable number. Also, it may be worth exploring the possibility and effectiveness of incorporating the full ASAF Fisher matrix in the analysis, especially for frequencies below 100~Hz where the pixel correlations may affect the upper limits.

The ASAF maps and Fisher information matrices were also appropriately integrated over frequency, for different choices of spectral models of the background, to recover the previously published search results for the broadband isotropic and anisotropic stochastic backgrounds. The ASAF results could be useful for many other future studies.

\section*{Acknowledgements}
This material is based upon work supported by NSF's LIGO Laboratory which is a major facility
fully funded by the National Science Foundation.
The authors also gratefully acknowledge the support of
the Science and Technology Facilities Council (STFC) of the
United Kingdom, the Max-Planck-Society (MPS), and the State of
Niedersachsen/Germany for support of the construction of Advanced LIGO 
and construction and operation of the GEO600 detector. 
Additional support for Advanced LIGO was provided by the Australian Research Council.
The authors gratefully acknowledge the Italian Istituto Nazionale di Fisica Nucleare (INFN),  
the French Centre National de la Recherche Scientifique (CNRS) and
the Netherlands Organization for Scientific Research (NWO), 
for the construction and operation of the Virgo detector
and the creation and support of the EGO consortium. 
The authors also gratefully acknowledge research support from these agencies as well as by 
the Council of Scientific and Industrial Research of India, 
the Department of Science and Technology, India,
the Science \& Engineering Research Board (SERB), India,
the Ministry of Human Resource Development, India,
the Spanish Agencia Estatal de Investigaci\'on (AEI),
the Spanish Ministerio de Ciencia e Innovaci\'on and Ministerio de Universidades,
the Conselleria de Fons Europeus, Universitat i Cultura and the Direcci\'o General de Pol\'{\i}tica Universitaria i Recerca del Govern de les Illes Balears,
the Conselleria d'Innovaci\'o, Universitats, Ci\`encia i Societat Digital de la Generalitat Valenciana and
the CERCA Programme Generalitat de Catalunya, Spain,
the National Science Centre of Poland and the European Union --- European Regional Development Fund; Foundation for Polish Science (FNP),
the Swiss National Science Foundation (SNSF),
the Russian Foundation for Basic Research, 
the Russian Science Foundation,
the European Commission,
the European Social Funds (ESF),
the European Regional Development Funds (ERDF),
the Royal Society, 
the Scottish Funding Council, 
the Scottish Universities Physics Alliance, 
the Hungarian Scientific Research Fund (OTKA),
the French Lyon Institute of Origins (LIO),
the Belgian Fonds de la Recherche Scientifique (FRS-FNRS), 
Actions de Recherche Concert\'ees (ARC) and
Fonds Wetenschappelijk Onderzoek --- Vlaanderen (FWO), Belgium,
the Paris \^{I}le-de-France Region, 
the National Research, Development and Innovation Office Hungary (NKFIH), 
the National Research Foundation of Korea,
the Natural Science and Engineering Research Council Canada,
Canadian Foundation for Innovation (CFI),
the Brazilian Ministry of Science, Technology, and Innovations,
the International Center for Theoretical Physics South American Institute for Fundamental Research (ICTP-SAIFR), 
the Research Grants Council of Hong Kong,
the National Natural Science Foundation of China (NSFC),
the Leverhulme Trust, 
the Research Corporation, 
the Ministry of Science and Technology (MOST), Taiwan,
the United States Department of Energy,
and
the Kavli Foundation.
The authors gratefully acknowledge the support of the NSF, STFC, INFN and CNRS for provision of computational resources.
This work was supported by MEXT, JSPS Leading-edge Research Infrastructure Program, JSPS Grant-in-Aid for Specially Promoted Research 26000005, JSPS Grant-in-Aid for Scientific Research on Innovative Areas 2905: JP17H06358, JP17H06361 and JP17H06364, JSPS Core-to-Core Program A. Advanced Research Networks, JSPS Grant-in-Aid for Scientific Research (S) 17H06133 and 20H05639 , JSPS Grant-in-Aid for Transformative Research Areas (A) 20A203: JP20H05854, the joint research program of the Institute for Cosmic Ray Research, University of Tokyo, National Research Foundation (NRF) and Computing Infrastructure Project of KISTI-GSDC in Korea, Academia Sinica (AS), AS Grid Center (ASGC) and the Ministry of Science and Technology (MoST) in Taiwan under grants including AS-CDA-105-M06, Advanced Technology Center (ATC) of NAOJ, Mechanical Engineering Center of KEK.
We would like to thank all of the essential workers who put their health at risk during the COVID-19 pandemic, without whom we would not have been able to complete this work.

\bibliography{ref_ASAF}

\end{document}

%% file: LSC-Virgo-KAGRA-Authors-Feb-2021-prd.tex

\author{R.~Abbott}
\affiliation{LIGO Laboratory, California Institute of Technology, Pasadena, CA 91125, USA}
\author{T.~D.~Abbott}
\affiliation{Louisiana State University, Baton Rouge, LA 70803, USA}
\author{F.~Acernese}
\affiliation{Dipartimento di Farmacia, Universit\`a di Salerno, I-84084 Fisciano, Salerno, Italy}
\affiliation{INFN, Sezione di Napoli, Complesso Universitario di Monte S. Angelo, I-80126 Napoli, Italy}
\author{K.~Ackley}
\affiliation{OzGrav, School of Physics \& Astronomy, Monash University, Clayton 3800, Victoria, Australia}
\author{C.~Adams}
\affiliation{LIGO Livingston Observatory, Livingston, LA 70754, USA}
\author{N.~Adhikari}
\affiliation{University of Wisconsin-Milwaukee, Milwaukee, WI 53201, USA}
\author{R.~X.~Adhikari}
\affiliation{LIGO Laboratory, California Institute of Technology, Pasadena, CA 91125, USA}
\author{V.~B.~Adya}
\affiliation{OzGrav, Australian National University, Canberra, Australian Capital Territory 0200, Australia}
\author{C.~Affeldt}
\affiliation{Max Planck Institute for Gravitational Physics (Albert Einstein Institute), D-30167 Hannover, Germany}
\affiliation{Leibniz Universit\"at Hannover, D-30167 Hannover, Germany}
\author{D.~Agarwal}
\affiliation{Inter-University Centre for Astronomy and Astrophysics, Pune 411007, India}
% ...

\collaboration{The LIGO Scientific Collaboration, the Virgo Collaboration, and the KAGRA Collaboration}